\documentclass[journal,12pt,onecolumn,draftclsnofoot,]{IEEEtran}
\usepackage{graphicx}  % Written by David Carlisle and Sebastian Raht
\usepackage{url}       % Written by Donald Arseneau
\usepackage{amsmath}   % From the American Mathematical Society
\usepackage{cite}
\usepackage{amsfonts,amssymb}
\usepackage{cases}
\usepackage{amsmath,amsfonts}
\usepackage{algorithmic}
\usepackage{algorithm}
\usepackage{array}
\usepackage[caption=false,font=normalsize,labelfont=sf,textfont=sf]{subfig}
\usepackage{textcomp}
\usepackage{stfloats}
\usepackage{url}
\usepackage{verbatim}
\usepackage{graphicx}
\usepackage{cite}
\def\BibTeX{{\rm B\kern-.05em{\sc i\kern-.025em b}\kern-.08em
    T\kern-.1667em\lower.7ex\hbox{E}\kern-.125emX}}
\usepackage{balance}
\begin{document}
%
% paper title
% can use linebreaks \\ within to get better formatting as desired
%\title{ Full-duplex One-way Amplify-and-forward Relay System Scheme for OFDM}
% 无邮箱开始
\title{A Joint Design  for Full-duplex OFDM AF Relay System with Precoded Short Guard Interval}

\author{Pu Yang, Xiang-Gen Xia,  Qingyue Qu, Han Wang and Yi Liu \\

\thanks{ The work of P. Yang, Q. Qu,  H. Wang  and Y. Liu was supported
%in part by the National Key R\&D Program of China (Grant No.2021YFA0716500), and
in part by 111 Project  under Grant B08038. (Corresponding author: Yi Liu.)}
% <-this % stops a space
\thanks{P. Yang, Q. Qu,  H. Wang and Y. Liu are with the State Key Laboratory
of Integrated Service Network, Xidian University, Xi'an 710071, China
(e-mail:yangp@stu.xidian.edu.cn; 22011211002@stu.xidian.edu.cn; hwang0815@stu.xidian.edu.cn; yliu@xidian.edu.cn).}
\thanks{X.-G. Xia is with the Department of Electrical and Computer
Engineering, University of Delaware, Newark, DE 19716, USA (e-mail: xxia@ee.udel.edu).}}
%
% The paper headers
%\markboth{Journal of \LaTeX\ Class Files,~Vol.~14, No.~8, August~2021}%
%{Shell \MakeLowercase{\textit{et al.}}: A Sample Article Using IEEEtran.cls for IEEE Journals}

%\IEEEpubid{0000--0000/00\$00.00~\copyright~2021 IEEE}
%% Remember, if you use this you must call \IEEEpubidadjcol in the second
%% column for its text to clear the IEEEpubid mark.

\maketitle
\begin{abstract}
In-band full-duplex relay (FDR) has attracted much attention as an effective solution to improve the coverage and spectral efficiency in  wireless communication networks.
The basic problem for FDR transmission is how to
eliminate the inherent self-interference and re-use the residual
self-interference (RSI) at the relay to improve the end-to-end
performance.
Considering the RSI at the FDR, the overall equivalent channel can be modeled as an  infinite impulse response (IIR) channel.  For this  IIR channel, a joint design for  precoding, power gain control and equalization of  cooperative OFDM relay systems  is presented. Compared with the traditional OFDM systems, the length of the  guard interval   for the proposed design  can be distinctly reduced, thereby improving the spectral efficiency. By analyzing  the noise sources, this paper evaluates the signal to  noise ratio (SNR) of the proposed scheme and presents a  power gain control algorithm at the FDR. Compared with the existing schemes, the proposed scheme shows  a superior  bit error rate (BER) performance.
\end{abstract}


%\IEEEpeerreviewmaketitle
\section{Introduction}
\IEEEPARstart{A}s the ever-increasing demand for  the limited wireless resources,
%efficient use of the radio spectrum has become more urgent.
in-band full-duplex relay (FDR) has gained significant
attention  due to its potential for  improving spectral efficiency and network coverage. Recent progress achieved in self-interference cancellation (SIC) has made the implementation of full-duplex relay  possible. After   passive and active SIC, %~\cite{FullDuplexAntenna2022}~\cite{Inherent2018},
there is still residual self-interference (RSI) existed at the relay.
% It  is  a trade-off between the system complexity and the ability of SIC at the FDR.
 An important issue for FDR networks is how to model and reuse the RSI to improve the overall system performance.

The formulation of the RSI is studied in~\cite{Generalized2019,PerformanceofFullDuplexAFRelaying6832455}.
As for the usage of RSI, paper~\cite{FiniteOrder2020} indicates that  the RSI at the relay is, in fact, a delayed version of the desired signal, and the FDR can be  considered as  an infinite impulse response (IIR) filter.
%in can  be exploited in the transceiver optimization.
%The FD relay is  then considered as  an infinite impulse response (IIR) filter in ~\cite{FiniteOrder2020}, and finite-order filtering designs  for  source and FDR  are investigated. %In~\cite{2017Distributed}, the residual SI at the FDR is  re-used to realize  a  full cooperative diversity space-time code.
%Furthermore in~\cite{Rii2009Optimiz12} and~\cite{LiXiaofeng2018OptimalTraining}, considering  the %feedback of the
% RSI  at the relay,
%the  source-to-destination   channel  becomes an frequency-selective channel.
% In recent studies~\cite{FiniteOrder2020} and~\cite{Inter2022Relay},
%In summary, the recursive RSI at the FDR results in  the  end-to-end  channel becoming  an IIR channel, and corresponding transmission techniques should be studied to avoid the  inter-symbol-interference (ISI).
In~\cite{LiXiaofeng2018OptimalTraining,Yi2017SC13,Kim2012},   the source-to-destination   IIR channel is  approximated by   a finite impulse response (FIR)  channel by choosing an effective length $L$ of the channel impulse responses wherein most of the energy (e.g. 99\%).
As for the cyclic prefix (CP) added transmission  format, a block-based transmission with a guard interval (GI) length
larger than $L$ symbols can  basically   avoid the ISI. For example, with a CP length $L\!=\!16$, the scheme  proposed in~\cite{Yi2017SC13}, which uses the traditional frequency domain equalization, has a good performance. However  in some applications, the  impulse response of  the IIR channel reduces slowly, and thus cannot be well approximated by a short FIR channel. Recently in~\cite{ANewxia}, a new OFDM system for an IIR channel is presented.  By a special design for the  GI,  an IIR channel can be  converted to ISI free subchannels at the receiver.

In this paper, based on the end-to-end equivalent IIR model for  FDR transmission  and the newly proposed OFDM system for IIR channels in~\cite{ANewxia},  we present a  joint design of the  precoding method at the source, power gain control algorithm at the  FDR, and a low complexity receiver  at the  destination for a cooperative OFDM system.
% The length of the guard interval (GI) in this design is extremely short so the  spectral efficiency  of the system is further improved.  A theoretical analysis and simulation results  verify  that the  proposed system can improve the end-to-end performance  under strong RSI.
The remainder of this letter is organized as follows. In Section~\ref{sec:sys2},  the system model is  presented. In Section~\ref{sec:Prec3}, we consider the precoding method   and equalization method for  the IIR channel. In Section~\ref{sec:Antenna}, the analysis of the SNR and  a power gain control method  are presented. Simulation results are given in Section~\ref{sec:simulation} and  the paper is concluded in Section~\ref{sec:con6}.
\section{ Equivalent IIR Channel  Model  }\label{sec:sys2}
%In this section, the characteristic of the RSI at FDR is studied and  equivalent channel models with and without direct link  are established.
%These  models use $z$-transform to give  a integrated description of the FDR transmission process where the  RSI signals are equalized as IIR channels.
The system consists of a source S, a  destination D and an  amplify-and-forward (AF)  FDR  R.  In time slot $n,n\geq0$, S transmits $x_n$ and D receives $y_n$, while the FDR transmits $t_n$ and receives $r_n$ simultaneously.
We  assume the  point-to-point  link channels are  quasi-static Rayleigh fading channels.
$h_{sr}$,  $h_{rd}$ and $h_{sd}$  are the channel coefficients for  S-to-R channel, R-to-D channel and S-to-D channel,  respectively.  After the SIC process, the RSI channel can be modeled as a quasi-static Rayleigh fading channel~\cite{Yi2017SC13} of channel coefficient  $h_{rr}$.
%We  assume that the RSI  variance increases linearly with the transmission  power.
Assume $({n_R})_{n}\sim\mathcal{CN}(0,\sigma^{2}_{R})$ and $({n_D})_{n}\sim\mathcal{CN}(0,\sigma^{2}_{D})$  as the complex-valued  white Gaussian noise with mean 0 and variances $\sigma^{2}_{R}$ and $\sigma^{2}_{D}$  at relay R and destination D at time $n$, respectively.
The information  symbols at the source
\begin{equation}		
		{{\mathbf{X}}}=[X_0, X_1, ..., X_{N-1}]^T,
\label{eq2227}
\end{equation}
are assumed to be statistically independent, identically distributed (i.i.d.) random variables.
After the $N$-point IFFT, the source  transmits signals by block $\mathbf{x}$,
\begin{eqnarray}
\begin{split}
{{\mathbf{x}}} = {\text{IFFT}}({{\mathbf{X}}}) = {\left[ {x_0,x_1, \cdots,x_{N - 1}} \right]^T}.
\label{equ2}
\end{split}
\end{eqnarray}
For large $N$, the  samples $x_n$ of    OFDM symbols are asymptotically Gaussian and i.i.d.~\cite{cho2010mimo}.
%Gaussian distributed, and they are uncorrelated thus statistically independent~\cite{distribution905885}.
%If $N$ is large, the  samples of OFDM symbols    are  asymptotically Gaussian distributed, and they are independent~\cite{cho2010mimo}~\cite{distribution905885}.
The received  signal at the FDR  at time $n$ is
\begin{eqnarray}
\begin{split}
r_n&=h_{sr} x_n + h_{rr} t_n +(n_{R})_n.
\label{eq1new}
\end{split}
\end{eqnarray}Suppose the amplification factor of the relay is $\beta$. Then, the power gain is  $\beta^2$. Following~\cite{LiXiaofeng2018OptimalTraining} and~\cite{Kim2012},  assume there is one symbol processing delay for the relay to forward its received symbols. The signal transmitted from R is
%\begin{equation}
%t(i)=
%\begin{cases}
%0& \text{i=0}\\
%\beta r(i-1)& \text{i$\geq$1.}
%\end{cases}
%\label{eq2}
%\end{equation}
%{\setlength\abovedisplayskip{3pt}
%\setlength\belowdisplayskip{3pt}
%\begin{equation}		
%		\resizebox{0.4\hsize}{!}{$\begin{aligned}
%		 t_n=
%\begin{cases}
%0,& n=0,\\
%\beta r_{n-1},& n\geq1.
%\end{cases}
%\label{eq2}
%\end{aligned}$}
%\end{equation}}
\begin{equation}		
		 t_n=\beta r_{n-1}.
\label{eq2}
\end{equation}From (\ref{eq1new}) and  (\ref{eq2}), we obtain
%{\setlength\abovedisplayskip{3pt}
%\setlength\belowdisplayskip{3pt}
%\begin{eqnarray}
%\begin{split}
%t_n=\beta h_{rr} t_{n-1}+\beta\left[h_{sr} x_{n-1}+(n_{R})_{n-1}\right].
%\label{eq3}
%\end{split}
%\end{eqnarray}}
\begin{align}
 \hspace{-0.15cm} {t_n} &=\beta  {{h_{sr}}{x_{n - 1}} +  \beta {{({n_R})}_{n - 1}}} +\beta {h_{rr}}{t_{n - 1}} {\normalsize\label{eq3}}\\
   &=\beta{h_{sr}} \sum\limits_{j = 1}^\infty  {{{\left( {\beta {h_{rr}}} \right)}^{j - 1}}} {x_{n - j}} + \beta \sum\limits_{j = 1}^\infty  {{{\left( {\beta {h_{rr}}} \right)}^{j - 1}}} {({{n}_R})_{n - j.}}
{\normalsize\label{eq1740}}
\end{align}Without the direct link, the  received  signal at  D at time $n$ is
%{\setlength\abovedisplayskip{3pt}
%\setlength\belowdisplayskip{3pt}
%\begin{align}
%y_n=h_{rd} t_n+(n_{D})_n.
%\label{eqa630}
%\end{align}}
%{\setlength\abovedisplayskip{3pt}
%\setlength\belowdisplayskip{3pt}
%\begin{align}
% {y_n} &= {h_{rd}}{t_n} + {({n_D})_n}
%   = \beta {h_{rd}}{h_{sr}}\sum\limits_{j = 1}^\infty  {{{\left( {\beta {h_{rr}}} \right)}^{j - 1}}} {x_{n - j}} \notag \\
%   &+\beta {h_{rd}}\sum\limits_{j = 1}^\infty  {{{\left( {\beta {h_{rr}}} \right)}^{j - 1}}} {({n_R})_{n - j}} + {({n_D})_n}.
%{\normalsize\label{eq630}}
%\end{align}}
\begin{equation}		{\begin{aligned}
		{y_n}=&{h_{rd}}{t_n} + {({n_D})_n} \\
=& \beta {h_{rd}}{h_{sr}}\sum\limits_{j = 1}^\infty  {{{\left( {\beta {h_{rr}}} \right)}^{j - 1}}} {x_{n - j}} \\
  &+\beta {h_{rd}}\sum\limits_{j = 1}^\infty  {{{\left( {\beta {h_{rr}}} \right)}^{j - 1}}} {({n_R})_{n - j}} + {({n_D})_n}.
\label{eq630}
\end{aligned}}
\end{equation}Take the  $z$-transform on both sides of  equation (\ref{eq3}):
\begin{align}
T(z) =  \beta [{h_{sr}}{z^{ - 1}}X(z) + {z^{ - 1}}{N_R}(z)] +\beta {h_{rr}}{z^{ - 1}}T(z).
\label{eqa6}
\end{align}Then, the system transfer  function between S and D is
\begin{equation}		
	\begin{aligned}
		 H(z) &= \frac{{{h_{rd}}T(z)}}{{X(z)}} \\
&= \frac{{\beta {h_{sr}}{h_{rd}}{z^{ - 1}}}}{{1 - \beta {h_{rr}}{z^{ - 1}}}} \\
&= \frac{1}{{\frac{1}{{\beta {h_{sr}}{h_{rd}}}} - \frac{{{h_{rr}}}}{{{h_{sr}}{h_{rd}}}}{z^{ - 1}}}}{z^{ - 1}},\quad \left| {\beta {h_{rr}}} \right| < \left| z \right| < \infty.
\label{eqa775}
\end{aligned}
\end{equation}Due to the fact that the overall time delay $z^{ - 1}$ does not  change the system performance,  we  use the transfer  function ${H_1}(z)$  to describe the equivalent channel without direct link:
\begin{figure}
%\setlength{\abovecaptionskip}{0.03cm}
%\vspace{-0.3cm}
\centering
\includegraphics[scale=0.25]{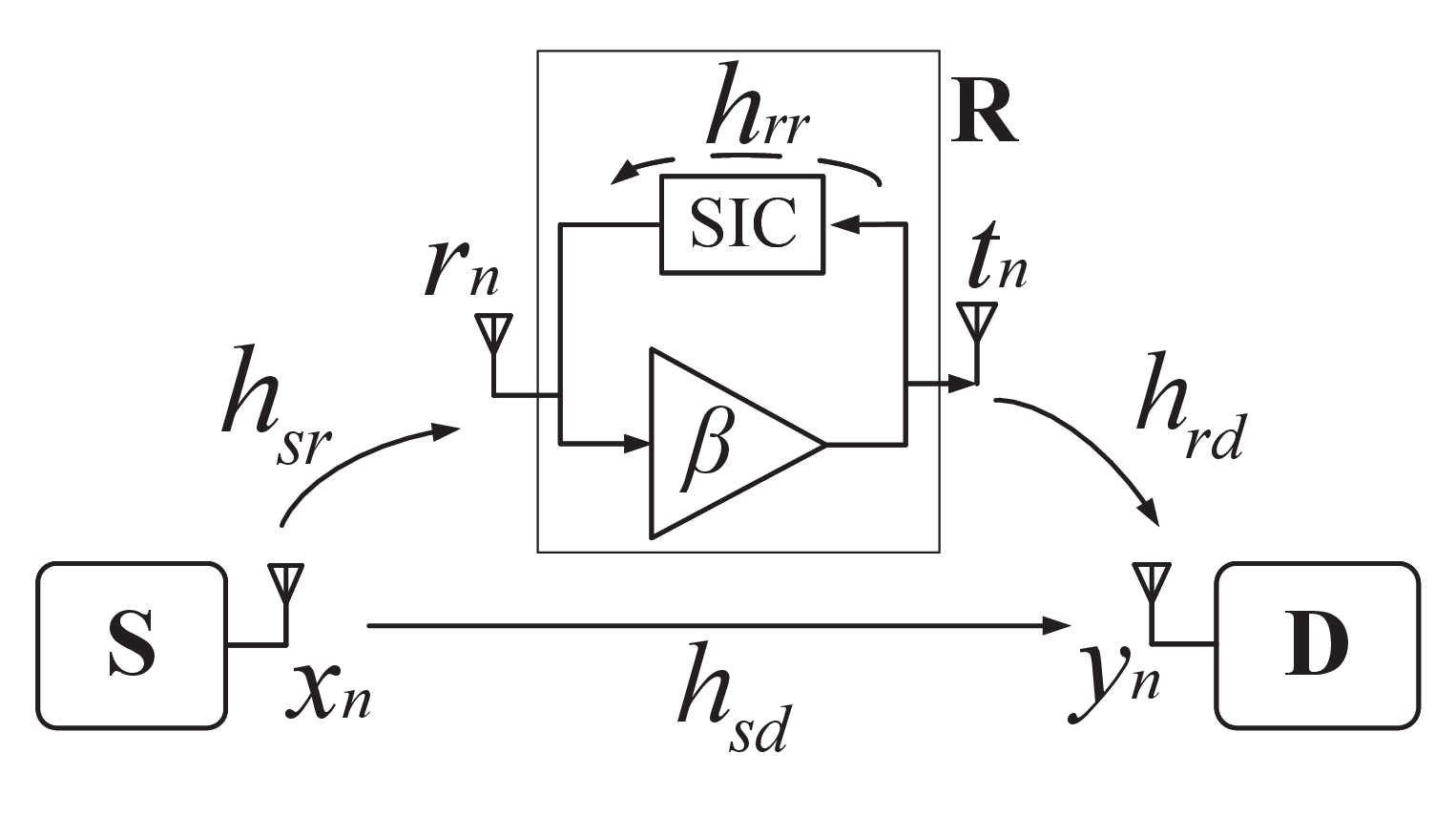}
\caption{Equivalent IIR channel  model.} \label{fig:zmodel}
%\vspace{-0.6cm}
\end{figure}

\begin{equation}	
		\begin{aligned}
  {H_1}(z) &= \frac{1}{{\frac{1}{{\beta {h_{sr}}{h_{rd}}}} - \frac{{{h_{rr}}}}{{{h_{sr}}{h_{rd}}}}{z^{ - 1}}}} \\
  &= \frac{1}{{A(z)}} = \frac{1}{{\sum\limits_{k = 0}^1 {{a_k}{z^{ - k}}} }} , \\
  {a_0} &= \frac{1}{{\beta {h_{sr}}{h_{rd}}}},\quad {a_1} \!= \! - \frac{{ {h_{rr}}}}{{ {h_{sr}}{h_{rd}}}}. \\
\end{aligned}
\label{eq175}
\end{equation}One can see that ${H_1}(z)$ is a single pole IIR channel. To ensure the  stability of the system, the pole of $ {H_1}(z)$, ${\beta {h_{rr}}}$, should satisfy
  \begin{equation}	
	\left| {\beta {h_{rr}}} \right|<1.
\label{eq1759}
\end{equation}When  $|\beta {h_{rr}}|$ is close to  $1$, the  impulse response
of the equivalent   IIR  channel reduces slowly and thus  cannot be well approximated by a short FIR channel.

If  the  direct link is considered, the equivalent channel has the following mixed first-order IIR channel transfer  function ${H_2}(z)$:
%{\setlength\abovedisplayskip{4pt}
%\setlength\belowdisplayskip{4pt}
%\begin{equation}
%\resizebox{0.66\hsize}{!}{$\begin{aligned}		
%	\begin{gathered}
%  {H_2}(z) = \frac{{{h_{sd}} + (\beta {h_{sr}}{h_{rd}} - \beta {h_{rr}}{h_{sd}}){z^{ - 1}}}}{{1 - \beta {h_{rr}}{z^{ - 1}}}} \\
%  \begin{array}{*{20}{c}}
%  { = \frac{{B(z)}}{{A(z)}} = \frac{{\sum\limits_{k = 0}^1 {{b_k}{z^{ - k}}} }}{{\sum\limits_{k = 0}^1 {{a_k}{z^{ - k}}} }},}&{\left| {\beta {h_{rr}}} \right|}
%\end{array} < \left| z \right| < \infty.\\
%\end{gathered}
%\label{eq12}
%\end{aligned}$}
%\end{equation}}
\begin{equation}	
	\begin{aligned}
  {H_2}(z) &= {h_{sd}} + \frac{{\beta {h_{sr}}{h_{rd}}{z^{ - 1}}}}{{1 - \beta {h_{rr}}{z^{ - 1}}}} \\
  &= \frac{{{h_{sd}} + (\beta {h_{sr}}{h_{rd}} - \beta {h_{rr}}{h_{sd}}){z^{ - 1}}}}{{1 - \beta {h_{rr}}{z^{ - 1}}}} \hfill \\
   &= \frac{{B(z)}}{{A(z)}} \\
  &= \frac{{\sum\limits_{k = 0}^1 {{b_k}{z^{ - k}}} }}{{\sum\limits_{k = 0}^1 {{a_k}{z^{ - k}}} }},\quad{\left| {\beta {h_{rr}}} \right|} < \left| z \right| < \infty .
\end{aligned}
\label{eq12}
\end{equation}
%We can see that due to the RSI at the relay, the overall  channel becomes a  frequency-selective fading IIR channel.
%The above  $z$-domain model gives  a complete representation of the channel impulse responses
% in AF FDR networks, which is the basis in the following design and analysis.
\section{ Precoding  and Equalization Method for Equivalent IIR Channel with Short GI  }\label{sec:Prec3}
Following the recently proposed OFDM  system for IIR channels  in~\cite{ANewxia},  a  precoding method and a corresponding frequency domain equalization method for the   above  equivalent IIR channel are  presented in this section. Due to the special structure of the channels in  (\ref{eq175}) and  (\ref{eq12}), i.e.,  order 1 IIR channels,  the designs have  simple and closed forms  in time domain. Furthermore, the noise terms can be analyzed well  as we shall see in the next section.
\setlength{\parskip}{0pt}

The  main goal of the precoding  is to obtain a sequence of standard CP structure after the equivalent IIR channel. Then, the transmitted signal from the source node   can be solved by frequency domain equalization without ISI.
%When the coefficients of the channels are known at the transmitter, the GI of an OFDM symbol can be designed at the transmitter by the proposed method.
Following~\cite{ANewxia}, the length of the GI should be  the same or larger than the orders of
polynomials $A(z)$ and $B(z)$. For the cases of ${H_1}(z)$ and ${H_2}(z)$ in (\ref{eq175}) and (\ref{eq12}), respectively, we can set  the length  of the GI as  $L=1$. We use $\mathbf{x}$ as the whole transmitted  sequence with  GI insertion, and $\mathbf{x}^i$ as the $i$th block without GI. The transmitted  sequence  at the source node is
%{\setlength\abovedisplayskip{4pt}
%\setlength\belowdisplayskip{4pt}
% \begin{eqnarray}
%{\mathbf{x}} \!= \!{[...;{\bar x_{N - 1}^{i - 1}},x_0^{i - 1},\! \cdots\! ,x_{N - 1}^{i - 1}\!;\!{\bar x_{N - 1}^i},\!x_0^i, \!\cdots \!,x_{N - 1}^i\!;\!...]^T},
%\label{equ22x}
%\end{eqnarray}}
 \begin{eqnarray}
{\mathbf{x}} = {[...;\,{\bar x_{N - 1}^{i - 1}},x_0^{i - 1}, ... ,x_{N - 1}^{i - 1};\,{\bar x_{N - 1}^i},\underbrace {x_0^i, ... ,x_{N - 1}^i}_{{{\mathbf{x}}^i}};\,...]^T},
\label{equ22x}
\end{eqnarray}where $\bar x_{N-1}^i$ is the inserted GI symbol
that  will be specially designed later.
The  corresponding received sequence $\mathbf{y}$ at the destination is
%
%{\setlength\abovedisplayskip{5pt}
%\setlength\belowdisplayskip{5pt}
%\begin{eqnarray}
%{\mathbf{y}} = {[...;\bar y_{N - 1}^{i - 1},y_0^{i - 1}, \cdots ,y_{N - 1}^{i - 1};\bar y_{N - 1}^i,y_0^i, \cdots ,y_{N - 1}^i;...]^T},
%\label{equ22y}
%\end{eqnarray}}
\begin{eqnarray}
{\mathbf{y}}={[{...;\,\bar y_{N - 1}^{i - 1},y_0^{i - 1},...,y_{N - 1}^{i - 1};\,\bar y_{N - 1}^i,\underbrace {y_0^i, ... ,y_{N - 1}^i}_{{{\mathbf{y}}^i}};\,...} ]^T},
\label{equ22y}
\end{eqnarray}where $\mathbf{y}^i$ is the $i$th received block without GI.
Let
\begin{equation}		
	\begin{gathered}
  {{\mathbf{Y}}^i} = {[Y_k^i]}_{0\le k\le {N-1}}\!= \!{\text{FFT(}}{{\mathbf{y}}^i}),\\ {{\mathbf{X}}^i} = [X_k^i]_{0\le k\le {N-1}}\! = \!{\text{FFT(}}{{\mathbf{x}}^i}), \\
  {\mathbf{A}} = [{A_k}]_{0\le k\le {N-1}} \!=\! {\text{FFT(}}{\mathbf{a}}),\quad{\mathbf{a}} \!= \! [{a_0},{a_1},0,...,0],
\end{gathered}
\label{equ517}
\end{equation}where FFT is the $N$-point FFT. We first consider the precoding method for the IIR equivalent channel without direct link,  ${H_1}(z)$.  The  goal is to design the GI $\bar x_{N - 1}^i$ to ensure  that the received  signals at the destination  satisfy the CP structure and in the case of this letter, it is  $\bar y_{N - 1}^i = y_{N - 1}^i$.
Let $X(z)$ and  $Y(z)$ be the  $z$-transforms of transmitted  sequence $\mathbf{x}$ and the corresponding received sequence $\mathbf{y}$. Then we have
\begin{equation}		
\begin{aligned}
  X(z) &= \frac{1}{{{H_1}(z)}}Y(z)\\
  &= A(z)Y(z) \\
  &= {a_0}Y(z) + {a_1}{z^{ - 1}}Y(z) \\
  \leftrightarrow {x_n} & = {a_0}{y_n} + {a_1}{y_{n - 1}}\\
   \Rightarrow {y_n} &= \frac{{{x_n} - {a_1}{y_{n - 1}}}}{{{a_0}}}.  \\
\end{aligned}
\label{xinequ228}
\end{equation}Similar to the conventional OFDM, we have
\begin{equation}
{X}_k^i= {A_k}Y_k^i \quad \!\!\! {\text{and}}\!\!\! \quad Y_k^i = \frac{{X_k^i}}{{{A_k}}} {\text{ ,}}\quad 0 \leqslant k \leqslant N - 1.
\label{equ1235}
\end{equation}Assume $A(z)$ is known at the source. According to (\ref{equ517}) and (\ref{equ1235}), $\mathbf{y}^i$ can be solved for given ${X}_k^i$ at the source. Finally, the GI at the source can be designed as
\begin{equation}		
	\bar x_{N - 1}^i = \left\{ {\begin{array}{*{20}{l}}
  {{a_0}y_{N - 1}^i,  \quad \quad \quad \quad \quad  \quad \! \! \!i  = 1,} \\
  {{a_0}y_{N - 1}^i + {a_1}y_{N - 1}^{i - 1}, \quad i > 1},
\end{array}} \right.
\label{equ100}
\end{equation}where the term for  $i=1$ is because the $0$th block is all 0 in the initialization.  By  this  precoding method,  the received sequence $\mathbf{y}$ at the destination  has a standard CP structure without the consideration of the noise. Thus, similar to  the traditional OFDM, the frequency domain equalized signal
  $\hat{{X}}^i_k$ at the $k$th  subcarrier is
\begin{equation}
\hat {X}_k^i= {A_k}Y_k^i{\text{,}}\quad 0 \leqslant k \leqslant N - 1.
\label{equ123}
\end{equation}
%Denote  the $N$-point IFFT of  $\hat{{X}}^i_k$, $0 \!\! \leqslant \! k \! \leqslant  \!\!  N - 1$, as
%${\left( {{\mathbf{\hat x}}^i} \right)^T} \!\!= \!\!{\left[ {{{\hat x}^i}_0,{{\hat x}^i}_1, \cdots ,{{\hat x}^i}_{N - 1}})\right]^T}$.
Let
\begin{equation}
{\text{IFFT(}}{[\hat X_k^i]_{0 \leqslant k \leqslant N - 1}}) = {\left( {{{{\mathbf{\hat x}}}^i}} \right)^T} = {\left[ {{{\hat x}^i}_0,{{\hat x}^i}_1, \cdots ,{{\hat x}^i}_{N - 1})} \right]^T},
\label{equ123}
\end{equation}
where  IFFT is the $N$-point IFFT.
%Due to the  circular convolution property of  the DFT, the frequency domain equalization in (\ref{equ123}) actually performs a circular convolution on the received sequence ${{{\mathbf{y}}^i}}$ with ${\mathbf{a}=[a_0,a_1]}$ in time domain.
Then,  the equalized signals in time domain can be expressed as
\begin{eqnarray}
\begin{split}
\left\{ {\begin{array}{*{20}{l}}
  {\hat x_0^i = {a_0}y_0^i + {a_1}y_{N - 1}^i}, \\
  {\hat x_n^i = {a_0}y_n^i + {a_1}y_{n - 1}^i,\quad 1 \leqslant n \leqslant N - 1}.
\end{array}} \right.
\label{equ22281}
\end{split}
\end{eqnarray}Due to the design  that  $\bar y_{N - 1}^i\!=\!y_{N - 1}^i$, we can see that (\ref{equ22281}) returns to the original signal $x_n^i$ without the consideration of noise.
%From  (\ref{xinequ228}), the actual values of the transmitted  ${{{\mathbf{x}}^i}}$ from the source should meet the following equations:
%{\setlength\abovedisplayskip{4pt}
%\setlength\belowdisplayskip{4pt}
%\begin{eqnarray}
%\begin{split}
%\left\{ {\begin{array}{*{20}{l}}
%  { x_0^i = {a_0}y_0^i + {a_1} \bar y_{N - 1}^i} ,\\
%  { x_n^i = {a_0}y_n^i + {a_1}y_{n - 1}^i, \quad 1 \leqslant n \leqslant N - 1}.
%\end{array}} \right.
%\label{equ22282}
%\end{split}
%\end{eqnarray}}

 %one  in (\ref{equ22282}), i.e., the transmitted signal can be correctly solved without ISI.

As for the equivalent channel with direct link,  ${H_2}(z)$, it is a mixed IIR channel. Following (\ref{xinequ228}), the precoding is the same as the pure IIR channel with polynomial $A(z)$ as above and we then define  an intermediate $z$-domain response $C(z)$ as
\begin{equation}	
\begin{aligned}
Y(z)&={H_2}(z)X(z)\\
&=\frac{{B(z)}}{{A(z)}}X(z)=B(z)C(z){\text{, }}\\
C(z)\!&=\!\frac{{X(z)}}{{A(z)}}.
\end{aligned}
\label{equ222}
\end{equation}The   corresponding sequence  ${\mathbf{c}}$ in time domain  is
%{\setlength\abovedisplayskip{5pt}
%\setlength\belowdisplayskip{5pt}
%\begin{eqnarray}
%\begin{split}
%{{\mathbf{c}}^i} = {\left[ {{\mathbf{\bar c}}_{}^i,{{\mathbf{c}}^i}} \right]^T} = {\left[ {\bar c_{N - 1}^i,c_0^i, \cdots ,c_{N - 1}^i} \right]^T}.
%\label{equ222}
%\end{split}
%\end{eqnarray}}
\begin{eqnarray}
{\mathbf{c}}\!=\!{[\!{...;\,\bar c_{N - 1}^{i - 1},c_0^{i - 1},...,c_{N - 1}^{i - 1};\,\bar c_{N - 1}^i,\underbrace {c_0^i, ... ,c_{N - 1}^i}_{{{\mathbf{c}}^i}};\,...} ]^T},
\label{equ22y}
\end{eqnarray}By  the precoding  for the pure IIR channel above, we can  get a sequence $\mathbf{c}$ with  standard CP,  i.e., $\bar c_{N - 1}^i\!= \!c_{N - 1}^i$.
Let
\begin{equation}	
\begin{gathered}	
	{\mathbf{ C}}^i = [ C_k^i]_{0\le k\le {N-1}} = {\text{FFT(}}{{\mathbf{c}}^i}),\\
{\mathbf{B}} \!= \![{B_k}]_{0\le k\le {N-1}} = {\text{FFT(}}{{\mathbf{b}}}),\quad  {\mathbf{b}} \!= \![{b_0},{b_1},0,...,0].
\end{gathered}
\label{equ1801}
\end{equation}As $\mathbf{y}$ is the response of $\mathbf{c}$ with the  FIR channel $B(z)$,  ${C}_k^i$ can be solved by frequency domain  equalization without ISI:
%$\hat{C}_k^i = \frac{{Y_k^i}}{{B_k}}$
\begin{eqnarray}
\begin{split}
{C}_k^i = \frac{{Y_k^i}}{{B_k}}.
\label{equ222}
\end{split}
\end{eqnarray}Finally, after a  two-step  frequency domain  equalization
\begin{eqnarray}
\begin{split}
\hat{X}_k^i = {A_k}{C}_k^i = \frac{{{A_k}Y_k^i}}{{B_k}},\quad 0\leq k\leq N-1,
\label{equ222}
\end{split}
\end{eqnarray}we can obtain the equalized signal $\hat{X}_k^i$ without ISI.
\section{Power Gain Control Algorithm at the Relay}\label{sec:Antenna}
By the precoding method and the OFDM approach  proposed in Section~\ref{sec:Prec3}, ISI free signals can be obtained at the destination. However, considering that the additive noise at the relay is also amplified as shown in (\ref{eq2})  during the transmission, improper  power gain at the FDR may  cause performance degradation.
In this section,  the noise during  the FDR transmission and the equalization process are analyzed in detail. An optimal power gain control algorithm  based on maximum SNR is presented.

The additive noise  $(n_{R})_n$ in (\ref{eq1new}) at the relay   and the additive noise $(n_{D})_n$ in (\ref{eq630}) at the destination  are the  main noise sources during the transmission.
The additive noise at the destination  after the  frequency domain  equalization is filtered by $\frac{1}{{H_1^{}(z)}}$  and its mean power can be expressed as
%\begin{equation}		
%		\resizebox{0.85\hsize}{!}{$\begin{aligned}
%	\begin{split}
%\begin{gathered}
%  {N_D} = \frac{1}{{2\pi j}}\oint_l {\sigma _D^2\frac{{(1 - \beta {h_{rr}}{z^{ - 1}})(1 - \beta {h_{rr}}z)}}{{{\beta ^2}{{\left| {{h_{rd}}} \right|}^2}{{\left| {{h_{sr}}} \right|}^2}}}} {z^{ - 1}}{\text{d}}z
%   = \frac{{1 + {{\left| {\beta {h_{rr}}} \right|}^2}}}{{{\beta ^2}{{\left| {{h_{sr}}} \right|}^2}{{\left| {{h_{rd}}} \right|}^2}}}\sigma _{\text{D}}^2.
%\end{gathered}
%\end{split}
%\end{aligned}$}
%\end{equation}
% 形式1
%{\setlength\abovedisplayskip{4pt}
%\setlength\belowdisplayskip{0pt}
%\begin{equation}		
%		\resizebox{0.85\hsize}{!}{$\begin{aligned}
%	\begin{gathered}
%  {P_D} = \frac{1}{{2\pi j}}\oint_l {\sigma _D^2\frac{1}{{H_1^{}(z)}}} \frac{1}{{H_1^{}({z^{ - 1}})}}{z^{ - 1}}{\text{d}}z \hfill \\
%   = \frac{1}{{2\pi j}}\oint_l {\sigma _D^2\frac{{(1 - \beta {h_{rr}}{z^{ - 1}})(1 - \beta {h_{rr}}z)}}{{{\beta ^2}{{\left| {{h_{rd}}} \right|}^2}{{\left| {{h_{sr}}} \right|}^2}}}} {z^{ - 1}}{\text{d}}z = \frac{{1 + {{\left| {\beta {h_{rr}}} \right|}^2}}}{{{\beta ^2}{{\left| {{h_{sr}}} \right|}^2}{{\left| {{h_{rd}}} \right|}^2}}}\sigma _{\text{D}}^2. \hfill \\
%\end{gathered}
%\end{aligned}$}
%\end{equation}
%}
% 形式2
\begin{equation}	
	\begin{aligned}
	{P_D}& = \frac{1}{{2\pi }}{\int_{ - \pi }^\pi  {\left| {\frac{1}{{H_1^{}({e^{j\omega }})}}} \right|} ^2}\sigma _D^2{\text{d}}\omega\\
& = \frac{{1 + {{\left| {\beta {h_{rr}}} \right|}^2}}}{{{\beta ^2}{{\left| {{h_{sr}}} \right|}^2}{{\left| {{h_{rd}}} \right|}^2}}}\sigma _{D}^2.{\text{ }}
 \end{aligned}
\end{equation}Next, the  noise part   caused by  the  additive  noise at the  relay is studied.
% Firstly, we define the notations of the noise terms  in different stages.
  Let $(n_{R})_n^i$ denote  the additive noise received by  the relay in the ${\it i}$th block at the $n$th time slot without the GI similar to $x_n^i$. $({{\bar n}_R})_{N - 1}^i$ denotes   the additive noise at the relay at  the  GI position of the $i$th block. The  $z$-domain  IIR channel of the relay  to the destination is denoted as $H_{rd}(z)$ that is
\begin{equation}		
 {H_{rd}}(z)\! = \!\frac{{\beta {h_{rd}}}}{{1 - \beta {h_{rr}}{z^{ - 1}}}} \!= \! \frac{{{H_1}(z)}}{{{h_{sr}}}}, \quad {\left| {\beta {h_{rr}}} \right|} \!\! <\!\! \left| z \right|\!\! < \infty .
\label{equ2227}
\end{equation}Let $({n_{R\_y}})_n^i$ represent the received noise at the destination  generated by $(n_{R})_n^i$ from the relay node  through the channel  ${H_{rd}}(z)$.
%Similar to (\ref{equ22282}) and
Considering the channel in (\ref{equ2227}), the following  holds
% {\setlength\abovedisplayskip{4pt}
%\setlength\belowdisplayskip{4pt}
% \begin{equation}		\resizebox{1\hsize}{!}{$\begin{aligned}
%	\left\{ {\begin{array}{*{20}{l}}
%  {({n_R})_0^i = {h_{sr}}{a_0}({n_{R\_y}})_0^i + {h_{sr}}{a_1}({{\bar n}_{R\_y}})_{N - 1}^i,} \\
%  {({n_R})_n^i = {h_{sr}}{a_0}({n_{R\_y}})_n^i + {h_{sr}}{a_1}({n_{R\_y}})_{n - 1}^i,1 \leqslant n \leqslant N - 1.}
%\end{array}} \right.
%\label{equ2244}
%\end{aligned}$}
%\end{equation}}
\begin{numcases}{}
 ({n_R})_0^i ={h_{sr}}({a_0}({n_{R\_y}})_0^i  + {a_1}({{\bar n}_{R\_y}})_{N - 1}^i), \label{eqsystem1} \\
({n_R})_n^i = {h_{sr}}({a_0}({n_{R\_y}})_n^i  +  {a_1}({n_{R\_y}})_{n - 1}^i),\quad 1 \leqslant n \leqslant N - 1, \label{eqsystem2}
\hspace{0.60cm}
\end{numcases}where the term $({{\bar n}_{R\_y}})_{N - 1}^i$  is the noise carried over  from the relay and  received by the destination  at the $i$th GI position  and will be specified later.
Note that
\begin{equation}	
{({\bar n}_{R\_y}})_{N - 1}^i   \ne {({n}_{R\_y}})_{N - 1}^i.
\label{eq1832}
 \end{equation}	
 This is because the GI design in  (\ref{equ100}) only uses the  transmitted signal $x_n$ but not the  noise that is unknown.
%The expression of ${({\bar n}_{R\_y}})_{N - 1}^i $ and ${({n}_{R\_y}})_{N - 1}^i$ will be presented later.
After  the  equalization, the noise caused by  $(n_{R})_n^i$ from the relay becomes $(\hat n_R)_n^i$. The frequency domain  equalization at the destination  performs  a circular convolution on the received noise, $({n_{R\_y}})_n^i$.   Similar to  (\ref{equ22281}), the noise after the equalization can be expressed as
%{\setlength\abovedisplayskip{4pt}
%\setlength\belowdisplayskip{4pt}
%\begin{eqnarray}
%\begin{split}
%\left\{ {\begin{array}{*{20}{l}}
%  {(\hat n_R)_0^i = {a_0}({n_{R\_y}})_0^i + {a_1} ({n_{R\_y}})_{N-1}^i},\\
%  {(\hat n_R)_n^i= {a_0}({n_{R\_y}})_n^i + {a_1} ({n_{R\_y}})_{n-1}^i ,1 \leqslant n \leqslant N - 1}.
%\end{array}} \right.
%\label{equ2228}
%\end{split}
%\end{eqnarray}}
%{\setlength\abovedisplayskip{4pt}
%\setlength\belowdisplayskip{4pt}
% \begin{equation}		
%		\resizebox{0.9\hsize}{!}{$\begin{aligned}
%\begin{split}
%\left\{ {\begin{array}{*{20}{l}}
%  {(\hat n_R)_0^i = {a_0}({n_{R\_y}})_0^i + {a_1} ({n_{R\_y}})_{N-1}^i},\\
% {(\hat n_R)_n^i= {a_0}({n_{R\_y}})_n^i + {a_1} ({n_{R\_y}})_{n-1}^i ,1 \leqslant n \leqslant N - 1}.
%\end{array}} \right.
%\end{split}
%\end{aligned}$}
%\end{equation}}
\begin{numcases}{}
	(\hat n_R)_0^i  = {a_0}({n_{R\_y}})_0^i + {a_1} ({n_{R\_y}})_{N-1}^i, \label{eqsystem3} \\
(\hat n_R)_n^i= \!{a_0}({n_{R\_y}})_n^i + {a_1} ({n_{R\_y}})_{n-1}^i , \quad 1 \leqslant n \leqslant N - 1. \label{eqsystem4}
\end{numcases}From (\ref{eqsystem2}) and  (\ref{eqsystem4}), we can see that  the noise terms $(\hat n_R)_1^i,...,(\hat n_R)_{N-1}^i $   are in linear relation with the additive noises $({n_R})_{1}^i,... ,({n_R})_{N-1}^i$ at the relay:
%{\setlength\abovedisplayskip{4pt}
%\setlength\belowdisplayskip{4pt}
%\begin{eqnarray}
%  (\hat n_R)_n^i = \frac{({n_R})_{n}^i}{h_{sr}},1 \leqslant n \leqslant N - 1.
%\label{equ2071}
%\end{eqnarray}}
 \begin{equation}		
(\hat n_R)_n^i = \frac{({n_R})_{n}^i}{h_{sr}},\quad\text{for}\quad 1 \leqslant n \leqslant N - 1.
\label{equ2071}
\end{equation}So the mean power of the noise terms from $({\hat n_R})_{1}^i$  to $({\hat n_R})_{N-1}^i$ is
 \begin{equation}		
{P_{R1}} = \frac{{\sigma _R^2}}{{{{\left| {{h_{sr}}} \right|}^2}}}.
\label{equ1851}
\end{equation}For  $(\hat n_R)_0^i$ in (\ref{eqsystem3}), the  exact expressions of ${({\bar n}_{R\_y}})_{N - 1}^i$,  ${({n}_{R\_y}})_{N - 1}^i$ and ${({n}_{R\_y}})_{0}^i$  are presented below.
%The  exact expressions of ${({\bar n}_{R\_y}})_{N - 1}^i$,  ${({n}_{R\_y}})_{N - 1}^i$ and ${({n}_{R\_y}})_{0}^i$  are presented in the following in preparation for calculating $(\hat n_R)_0^i$.
From (\ref{eq630}), the  received noise at the destination caused by  the additive noise at the relay is:
\begin{equation}			
	{({{{n}}_{R\_y}})_n} = h_{rd}\beta \sum\limits_{j = 1}^\infty  {{{\left( {\beta {h_{rr}}} \right)}^{j - 1}}} {({{{n}}_R})_{n - j + 1.}}	
\label{equ2119}
\end{equation}For simplicity, we define that
\begin{equation}			
	h_j=h_{rd}\beta(\beta h_{rr})^{j-1}, j\geqslant 1.
\label{equ2119}
\end{equation}From (\ref{equ2119}), the expressions  of $ ({{n}_{R\_y}})_{n}^i$  are
%\begin{align}
%  \left\{ {\begin{array}{*{20}{l}}
% \!\! {({{\bar n}_{R\_y}})_{N - 1}^i \!=\! ... \!+\! {h_2}({n_R})_{N - 1}^{i - 1}\! + \! {h_1}({{\bar n}_R})_{N - 1}^i,} \\
% \!\! {({n_{R\_y}})_0^i \!=\! ... \!+ \!{h_3}({n_R})_{N - 1}^{i - 1} \!+ \!{h_2}({{\bar n}_R})_{N - 1}^i \!+\! {h_1}({n_R})_0^i,} \\
%\! \! \cdots  \\
%  \!\!{({n_{R\_y}})_{N - 1}^i \!=\! ... \!+\! {h_{N + 1}}({{\bar n}_R})_{N - 1}^i \!+ \! {h_N}({n_R})_0^i \!+ \!...\! + \!{h_1}({n_R})_{N - 1}^i,}
%\end{array}} \right.
%\label{equ1898}
%\end{align}
%\begin{equation}			
%	\left\{ {\begin{array}{*{20}{l}}
%  {({{\bar n}_{R\_y}})_{N - 1}^i \!=\! ... \!+\! {h_2}({n_R})_{N - 1}^{i - 1}\! + \! {h_1}({{\bar n}_R})_{N - 1}^i,} \\
% {({n_{R\_y}})_0^i \!=\! ... \!+ \!{h_3}({n_R})_{N - 1}^{i - 1} \!+ \!{h_2}({{\bar n}_R})_{N - 1}^i \!+\! {h_1}({n_R})_0^i,} \\
% \cdots  \\
% {({n_{R\_y}})_{N - 1}^i \!=\! ... \!+\! {h_{N + 1}}({{\bar n}_R})_{N - 1}^i }\!+ \\
%  {{h_N}({n_R})_0^i \!+ ...\! + \!{h_1}({n_R})_{N - 1}^i,}
%\end{array}} \right.
%\label{equ1898}
%\end{equation}
%\begin{numcases}{}
%	\!\! {({{\bar n}_{R\_y}})_{N - 1}^i \!=\! ... \!+\! {h_2}({n_R})_{N - 1}^{i - 1}\! + \! {h_1}({{\bar n}_R})_{N - 1}^i,}  \nonumber\\
%	\!\! {({n_{R\_y}})_0^i \!=\! ... \!+ \!{h_3}({n_R})_{N - 1}^{i - 1} \!+ \!{h_2}({{\bar n}_R})_{N - 1}^i \!+\! {h_1}({n_R})_0^i,} \nonumber\\
%\! \! \cdots  {\label{equ1898}}\\
% \!\!{({n_{R\_y}})_{N - 1}^i \!=\! ... \!+\! {h_{N + 1}}({{\bar n}_R})_{N - 1}^i \!+ \! {h_N}({n_R})_0^i \!+ \!...\! + \!{h_1}({n_R})_{N - 1}^i,}\nonumber
%\end{numcases}
\begin{numcases}{}
	{({{\bar n}_{R\_y}})_{N - 1}^i = ... + {h_2}({n_R})_{N - 1}^{i - 1} +  {h_1}({{\bar n}_R})_{N - 1}^i,}  \nonumber\\
	 {({n_{R\_y}})_0^i \,\,\quad= ... + {h_3}({n_R})_{N - 1}^{i - 1} + {h_2}({{\bar n}_R})_{N - 1}^i + {h_1}({n_R})_0^i,} \nonumber\\
 \quad\quad \cdots  {\label{equ1898}}\\
 \!{({n_{R\_y}})_{N - 1}^i = ... + {h_{N + 1}}({{\bar n}_R})_{N - 1}^i }\nonumber\\
 \quad\quad\quad\quad\quad\quad{+ {h_N}({n_R})_0^i + ... + {h_1}({n_R})_{N - 1}^i,}\nonumber
\end{numcases}where term $({{\bar n}_R})_{N - 1}^i$ is the additive noise received by  the relay at  the $i$th GI position. Since  $\left| {\beta {h_{rr}}} \right| < 1$, and  considering
the length of each frame, $N$, is large,  we have $|h_j| \approx 0$ for all $j>N$.
 From (\ref{equ1898}), we can see that  ${({\bar n}_{R\_y}})_{N - 1}^i $ and  ${({n}_{R\_y}})_{N - 1}^i$ are  approximately  independent, so  $(\hat n_R)_0^i$ cannot be expressed   the same form as the other noise terms in (\ref{equ2071}).
 From (\ref{equ2119}),
  %we can see that  ${({n}_{R\_y}})_{0}^i$ and ${({n}_{R\_y}})_{N - 1}^i$  still have  the Gaussian distribution. Considering $N$ is large enough,
 the mean power of ${({n}_{R\_y}})_{0}^i$ and ${({n}_{R\_y}})_{N - 1}^i$
can be expressed as
% 形式1
% \begin{equation}		
%		\resizebox{0.85\hsize}{!}{$\begin{aligned}
%	\begin{gathered}
%  {P_n} = \frac{1}{{2\pi j}}\oint_l {\sigma _R^2H_{rd}^{}(z)H_{rd}^{}({z^{ - 1}}){z^{ - 1}}{\text{d}}z}  \hfill \\
%   = \frac{1}{{2\pi j}}\oint_l {\sigma _R^2\frac{{{\beta ^2}{{\left| {{h_{rd}}} \right|}^2}}}{{(1 - \beta {h_{rr}}{z^{ - 1}})(1 - \beta {h_{rr}}z)}}} {z^{ - 1}}{\text{d}}z = \frac{{{\beta ^2}{{\left| {{h_{rd}}} \right|}^2}}}{{1 - {{\left| {\beta {h_{rr}}} \right|}^2}}}\sigma _R^2. \hfill \\
%\end{gathered}
%\label{equ3275}
%\end{aligned}$}
%\end{equation}
% 形式2
%{\setlength\abovedisplayskip{4pt}
%\setlength\belowdisplayskip{4pt}
% \begin{equation}		
%		\resizebox{0.65\hsize}{!}{$\begin{aligned}
%	{P_n} = \frac{1}{{2\pi }}\int_{ - \pi }^\pi  {{{\left| {H_{rd}^{}({e^{j\omega }})} \right|}^2}} \sigma _R^2{\text{d}}\omega  = \frac{{{\beta ^2}{{\left| {{h_{rd}}} \right|}^2}}}{{1 - {{\left| {\beta {h_{rr}}} \right|}^2}}}\sigma _R^2.{\text{ }}
%\label{equ3275}
%\end{aligned}$}
%\end{equation}}
%形式3
 \begin{equation}	
 \begin{aligned}	
P_n&= {\left| {{h_{rd}}} \right|^2}{\beta ^2}\sum\limits_{j = 1}^\infty  {{{\left| {\beta {h_{rr}}} \right|}^{2(j - 1)}}} \sigma _R^2\\
&= \frac{{{\beta ^2}{{\left| {{h_{rd}}} \right|}^2}}}{{1 - {{\left| {\beta {h_{rr}}} \right|}^2}}}\sigma _R^2.
\label{equ3275}
\end{aligned}
\end{equation}From (\ref{equ1898}) and considering that $|h_j| \approx 0$ for all $j>N$,  ${({n}_{R\_y}})_{0}^i$ and ${({n}_{R\_y}})_{N - 1}^i$ can be regarded as   independent.
Then, from  (\ref{eqsystem3}), the mean power of $(\hat n_R)_0^i$ in(\ref{eqsystem3}) is
 \begin{equation}
 \begin{aligned}
	{P_{R2}} &= ({\left| {{a_0}} \right|^2}{P_n} + {\left| {{a_1}} \right|^2}{P_n}) \\
&= \frac{{1 + {{\left| {\beta {h_{rr}}} \right|}^2}}}{{{{\left| {{h_{sr}}} \right|}^2}(1 - {{\left| {\beta {h_{rr}}} \right|}^2})}}\sigma _R^2.
\end{aligned}
\label{equ3295}
\end{equation}Thus, the mean power of the  noise terms $(\hat n_R)_0^i ,...,(\hat n_R)_{N-1}^i $  can be expressed as
%As the received signal is finally demodulated in frequency domain, the noise term  $(\hat n_R)_0^i$ may be evenly distributed across all the subcarriers.
 %From (\ref{equ3295}), if $\left| {\beta {h_{rr}}} \right|$ is close to 1, the power of ${P_{R2}}$ will grow large. Therefore, a  proper  power gain  control algorithm  is necessary at the relay.
\begin{equation}	
 \begin{aligned}	
	{P_R} &= \frac{{{P_{R2}}+(N - 1){P_{R1}}}}{N} \\
&= \frac{{(N - 1)\sigma _R^2}}{{N{{\left| {{h_{sr}}} \right|}^2}}} + \frac{{(1\! + {{\left| {\beta {h_{rr}}} \right|}^2})\sigma _R^2}}{{N{{\left| {{h_{sr}}} \right|}^2}(1 - {{\left| {\beta {h_{rr}}} \right|}^2})}}.
 \end{aligned}
\label{equ6429}
\end{equation}Next, the power of the useful signals after equalization is calculated. Considering the mean power of transmitted signals should be normalized to 1,  the power of the GI, $\bar x_{N - 1}^i$,  should be clarified.  From (\ref{eq630}), the expression of $y_{N - 1}^i$ is
\begin{equation}		
	y_{N - 1}^i = {h_{sr}}\left( {... + {h_{N + 1}}\bar x_{N - 1}^i + {h_N}x_0^i + ... + {h_1}x_{N - 1}^i} \right).
\label{equ1702}
\end{equation}Since $|h_j| \approx 0$ for all $j>N$, the impact of $\bar x_{N - 1}^i$ in (\ref{equ1702}) can be ignored.
%and the symbols are uncorrelated
%~\cite{Envelope2010}.
The mean power of $y_{N - 1}^i$  or   $y_{N - 1}^{i - 1}$ is:
\begin{equation}		
	%{P_y} = {\left| {{h_{sr}}} \right|^2}{\left| {{h_{rd}}} \right|^2}{\beta ^2}\sum\limits_{j = 1}^\infty  {{{\left| {\beta {h_{rr}}} \right|}^{2(j - 1)}}} {\left| {{x_{n- j+1}}} \right|^2} = \frac{{{\beta ^2}{{\left| {{h_{sr}}} \right|}^2}{{\left| {{h_{rd}}} \right|}^2}}}{{1 - {{\left| {\beta {h_{rr}}} \right|}^2}}}\sigma_x^2,
	\begin{aligned}
{P_y} &= {\left| {{h_{sr}}} \right|^2}{\left| {{h_{rd}}} \right|^2}{\beta ^2}\sum\limits_{j = 1}^\infty  {{{\left| {\beta {h_{rr}}} \right|}^{2(j - 1)}}} \sigma_x^2 \\
&= \frac{{{\beta ^2}{{\left| {{h_{sr}}} \right|}^2}{{\left| {{h_{rd}}} \right|}^2}}}{{1 - {{\left| {\beta {h_{rr}}} \right|}^2}}}\sigma_x^2,
\end{aligned}
\label{eq1800}
\end{equation}where $\sigma _x^2$ is the mean power of the transmitted signals  without GI.
 From (\ref{equ1235}), since $y_{N - 1}^i$  and  $y_{N - 1}^{i - 1}$  are   determined  by ${X}_k^i$ and ${X}_k^{i-1}$, respectively, they are  statistically independent. From (\ref{equ100}), the mean power of  $\bar x_{N - 1}^i$ is
\begin{equation}
\begin{aligned}		
	{P_{GI}} &= ({\left| {{a_0}} \right|^2}{P_y} + {\left| {{a_1}} \right|^2}{P_y})\\
 &= \frac{{1 + {{\left| {\beta {h_{rr}}} \right|}^2}}}{{1 - {{\left| {\beta {h_{rr}}} \right|}^2}}}\sigma _x^2.
\label{eq1671}
\end{aligned}
\end{equation}Define $\alpha$  as
  \begin{equation}		
	\alpha\triangleq |h_{rr}\beta|^{2}\in(0,1).
\label{eq1836}
\end{equation}The mean power of the transmitted signals is
  \begin{equation}
  \begin{aligned}		
	P_{av}&= \frac{{{P_{GI}} + N\sigma _x^2}}{{N + 1}}\\
& = \frac{{\frac{{1 + \alpha }}{{1 - \alpha }} + N}}{{N + 1}}\sigma _x^2.
\end{aligned}
\label{eq1836}
\end{equation}
%{\setlength\abovedisplayskip{4pt}
%\setlength\belowdisplayskip{4pt}
%\begin{equation}		
%		\resizebox{0.75\hsize}{!}{$\begin{aligned}
%	{P_{average{\text{ }}}} = \frac{{{P_{GI}} + N\sigma _x^2}}{{N + 1}} = 1 \Rightarrow  \sigma _x^2 = \frac{{N + 1}}{{\frac{{1 + \alpha }}{{(1 - \alpha )}} + N}}.
%\end{aligned}$}
%\label{eq2254}
%\end{equation}}
%{\setlength\abovedisplayskip{4pt}
%\setlength\belowdisplayskip{4pt}
%\begin{equation}		
%		\resizebox{0.7\hsize}{!}{$\begin{aligned}
%	{P_{average{\text{ }}}} = \frac{{{P_{GI}} + N\sigma _x^2}}{{N + 1}} = \frac{{\frac{{1 + \alpha }}{{(1 - \alpha )}} + N}}{{N + 1}}\sigma _x^2.
%\end{aligned}$}
%\label{eq2254}
%\end{equation}}
If the   mean power of the transmitted signals is
normalized to 1, then the mean power of the transmitted signals without GI is
\begin{equation}	
\sigma _x^2 = \frac{{N + 1}}{{\frac{{1 + \alpha }}{{1 - \alpha }} + N}}.
\label{eq1837}
\end{equation}Let
\begin{equation}	
{P_{R1}} = \frac{{\sigma _R^2}}{{{{\left| {{h_{sr}}} \right|}^2}}},\quad
\eta  = \frac{{{{\left| {{h_{rr}}} \right|}^2}}}{{{{\left| {{h_{sr}}} \right|}^2}{{\left| {{h_{rd}}} \right|}^2}}}\sigma _D^2.
\label{eq1837}
\end{equation}
Finally, the SNR after equalization  is
\begin{equation}		
	\begin{aligned}
  \gamma &= \frac{{\sigma _x^2}}{{{P_R} + {P_D}}}{\text{ }} \\
  &= \frac{{\frac{{N + 1}}{{\frac{{1 + \alpha }}{{1 - \alpha }} + N}}}}{{\frac{{(N - 1)\sigma _R^2}}{{N{{\left| {{h_{sr}}} \right|}^2}}} + \frac{{(1 + \alpha )\sigma _R^2}}{{N{{\left| {{h_{sr}}} \right|}^2}(1 - \alpha )}} + \frac{{{{\left| {{h_{rr}}} \right|}^2}(1 + \alpha )}}{{\alpha {{\left| {{h_{sr}}} \right|}^2}{{\left| {{h_{rd}}} \right|}^2}}}\sigma _{D}^2}} \hfill \\
 & = \frac{{N(N + 1){{(\alpha  - 1)}^2}\alpha }}{{((\alpha\!  -\! 1)N - \alpha  - 1)(\eta ({\alpha ^2} - 1) N+ {P_{R1}}\alpha ((\alpha  - 1)N - 2\alpha ))}}.
\end{aligned}
\label{eq1823}
\end{equation}
Since $\beta^2$  is in  proportional with  $\alpha$, the optimization  of the power gain at the relay can be written as
\begin{eqnarray}
\begin{split}
\alpha_{opt}=arg \mathop{max} \limits_{\alpha}{(\gamma)}.
\label{eq18}
\end{split}
\end{eqnarray}Although in the above, we only considered the case when there is no direct link between S and D, from our simulations in next section, we find that the above power gain control algorithm is still valid when the direct link is not too strong. The general direct link case will be under our future study.
%-0.000206373955189753 - 0.000151890045357046i

Next, we compare the proposed scheme with a straightforward  pre-filtering method.
%To distinguish the  pre-filtering method from the proposed scheme,
For the  pre-filtering method,
let  ${\mathbf{s}}$ represent the OFDM  sequence  with a standard CP structure   generated at the source:
 \begin{eqnarray}
{\mathbf{s}} = {[...;\,s_{N - 1}^{i - 1},s_0^{i - 1}, ... ,s_{N - 1}^{i - 1};\,s_{N - 1}^i,s_0^i, ... ,s_{N - 1}^i;\,...]^T}.
\label{equ1742}
\end{eqnarray}The reason why OFDM with one symbol CP is used here is because when there is a direct link, the FIR part $B(z)$ appears as shown before. In this case, the conventional OFDM is used. Let   $\sigma_s^2$ denote the mean power of the signal $s_{n}^{i}$.
%in (\ref{equ1742}).
%Since  the source node  knows the channel parameters $a_n$,
%then after  the equivalent IIR channel, the received sequence at the destination is also ISI free.
Let ${{\tilde s}_n}$ denote the  pre-filtered  signal by the FIR filter $A(z)$, and
 \begin{equation}		
	{{\tilde s}_n} = {a_0}{s_n} + {a_1}{s_{n - 1}}.
\label{eq1911}
\end{equation}After  the  IIR channel  ${H_1}(z)$, ${{\tilde s}_n}$ is converted to the original signal ${s_n}$.
%the received sequence at the destination is converted to the original signal.
 The mean power of the transmitted signal ${{\tilde s}_n}$ is
 \begin{equation}	
 \begin{aligned}	
	{P_f} &= \left( {{{\left| {{a_0}} \right|}^2} + {{\left| {{a_1}} \right|}^2}} \right)\sigma _s^2 \\& = \frac{{1 + {{\left| {\beta {h_{rr}}} \right|}^2}}}{{{\beta ^2}{{\left| {{h_{rd}}} \right|}^2}{{\left| {{h_{sr}}} \right|}^2}}}\sigma _s^2.
\end{aligned}
\label{eq1911}
\end{equation}If the transmit power
 %{\setlength\abovedisplayskip{4pt}
%\setlength\belowdisplayskip{4pt}
% \begin{equation}		
%		\resizebox{0.89\hsize}{!}{$\begin{aligned}
%{{\tilde x}_n} \!\!= \!\!\frac{{{a_0}{x_n} \!+ \!{a_1}{x_{n - 1}}}}{{\sqrt {{{\left| {{a_0}} \right|}^2} + {{\left| {{a_1}} \right|}^2}} }} = \sqrt {\frac{{{\beta ^2}{{\left| {{h_{rd}}} \right|}^2}{{\left| {{h_{sr}}} \right|}^2}}}{{1 + {{\left| {\beta {h_{rr}}} \right|}^2}}}} \left( {{a_0}{x_n}\!\! + \!\!{a_1}{x_{n - 1}}} \right).
%\label{eq2309}
%\end{aligned}$}
%\end{equation}}
%{\setlength\abovedisplayskip{4pt}
%\setlength\belowdisplayskip{4pt}
% \begin{equation}		
%		\resizebox{0.80\hsize}{!}{$\begin{aligned}
%{P_{filtered}} = \frac{1}{{2\pi }}\int_{ - \pi }^\pi  {{{\left| {A({e^{j\omega }})} \right|}^2}} \sigma _s^2{\text{d}}\omega  = \frac{{1 + {{\left| {\beta {h_{rr}}} \right|}^2}}}{{{\beta ^2}{{\left| {{h_{rd}}} \right|}^2}{{\left| {{h_{sr}}} \right|}^2}}}\sigma _s^2.
%\label{eq2309}
%\end{aligned}$}
%\end{equation}}
%${P_{filtered}}$
is normalized to 1,  we have
 \begin{equation}		
	\sigma _s^2\!=\!{\frac{{{\beta ^2}{{\left| {{h_{sr}}} \right|}^2}{{\left| {{h_{rd}}} \right|}^2}}}{{1 + {\beta ^2}{{\left| {{h_{rr}}} \right|}^2}}}}.
\label{eq1911}
\end{equation}The power of the received noise at the destination generated by the additive noise  from the relay, ${P_{R3}}$,  can   be  calculated by (\ref{equ3275}).  Then, the SNR of the  pre-filtering method at the receiver is
\begin{equation}	
\begin{aligned}	
{\gamma _{pre}} &= \frac{{{\sigma_s^2}}}{{{P_{R3}} + \sigma _D^2}}\\
&= \frac{{\frac{{{\beta ^2}{{\left| {{h_{sr}}} \right|}^2}{{\left| {{h_{rd}}} \right|}^2}}}{{1 + {\beta ^2}{{\left| {{h_{rr}}} \right|}^2}}}}}{{\frac{{{\beta ^2}{{\left| {{h_{rd}}} \right|}^2}\sigma _R^2}}{{1 - {{\left| {\beta {h_{rr}}} \right|}^2}}} + \sigma _D^2}}. \!\!
%=\!\! \frac{1}{{\frac{{(1 + \alpha )}}{{(1 - \alpha ){{\left| {{h_{sr}}} \right|}^2}}}\sigma _R^2 + \frac{{{{\left| {{h_{rr}}} \right|}^2}(1 + \alpha )}}{{\alpha {{\left| {{h_{sr}}} \right|}^2}{{\left| {{h_{rd}}} \right|}^2}}}\sigma _D^2.}}
\end{aligned}
\label{eq1824}
\end{equation}Define the difference  between the SNR for our proposed method,  $\gamma$ in (\ref{eq1823}), and  the SNR for  the straightforward method,  $\gamma _{pre}$ in (\ref{eq1824}), as   $\Delta $ :
\begin{equation}
%\begin{align}
%\begin{split}
\begin{aligned}	
\Delta & \triangleq \gamma  - {\gamma _{{\text{pre}}}}\\
 &= \frac{{2(1 - \alpha ){\alpha ^2}(\overbrace {{P_{R1}}\alpha (1 - \alpha ){N^2} + ({P_{R1}}({\alpha ^2} - \alpha ) + \eta ({\alpha ^2} - 1))N - {P_{R1}}({\alpha ^2} - \alpha )}^{\delta (\alpha )})}}{{\left( {\alpha  + 1} \right)\left( {(\alpha  - 1)N - \alpha  - 1} \right)\left( {\eta (1 - \alpha ) + {P_{R1}}\alpha } \right)\left( {\eta ({\alpha ^2} - 1)N + {P_{R1}}\alpha (\alpha  - 1)N - 2{P_{R1}}{\alpha ^2}} \right)}}.
\end{aligned}	
\label{eq1509}
%\end{split}
%\end{align}
\end{equation}When $0 <\alpha < 1$, the
denominator of $\Delta$ is positive. As shown in (\ref{eq1509}), define a quadratic  function $\delta(\alpha)$ that  has the same sign with $\Delta $.
%xia
One can see that since $0<\alpha<1$, the coefficient of $N^2$ in $\delta(\alpha)$ is positive. So, when $N$ is large, $\delta(\alpha)>0$, i.e., the SNR $\gamma$ after the equalization of our proposed precoding method is better than the SNR $\gamma_{pre}$ of the straightforward prefiltering method. From the simulation results in next section, $\gamma\!>\!\gamma_{pre}$ when $N\!=\!128$.

\section{Simulation Results}\label{sec:simulation}
%In this section, the performance of the proposed precoding method is  evaluated in different aspects. Set $SNR_R=SNR_D=SNR$ and the signal power at the source node is normalized. Suppose that  $\sigma_{R}^{2}=\sigma_{D}^{2}$. The constellation used for FD schemes at source S  is QPSK. The average power loss  between source to relay, relay to destination are normalized. The number of sub-carriers   is $N=128$ and the length of guard interval is $L = 1$. The SIC ability of the relay is set as 15dB so the power of RSI is -15dB, which can be easily realized by a ferrite circulator or a pair of separate antennas.
 Suppose that the transmitted signal power at  the source and   the channel  gains
  of $h_{sr}$ and $h_{rd}$ are normalized to 1. The noise powers at R and D are assumed the same, i.e., ${\sigma_{R}^{2}}\!\!=\!\!{\sigma_{D}^{2}}$. The channel SNRs at R and D are defined as
   \begin{equation}
   SNR_c = \frac{1}{{\sigma _R^2}} = \frac{1}{{\sigma _D^2}}.
   \end{equation}
  %Note $SNR_n$ only dominates the power of the additive noise  at the relay and the destination and is different from the SNR ($\gamma$) analysed in last section.
  The number of sub-carriers  is $N\!\!=\!\!128$. The length of CP or GI for the following schemes is set as 1.
 % The length of GI for the proposed scheme is $L \!\!=\!\!1$.
%  For the other schemes， the length of the CP is 1.
  The SIC ability of the relay is set as 15dB so the power of RSI is -15dB. The constellation  for FD schemes at the source  is QPSK. We compare the proposed  scheme with four  FD  schemes including  Wichman scheme~\cite{Rii2009Optimiz12}, SC-FDE scheme~\cite{Yi2017SC13}, the traditional OFDM frequency domain equalization scheme with a standard CP structure~\cite{cho2010mimo} and the straightforward  pre-filtering scheme mentioned in Section~\ref{sec:Antenna}. The scheme in~\cite{Rii2009Optimiz12} treats RSI as noise.
 % The direct link is ignored in~\cite{Yi2017SC13}.
   The traditional OFDM scheme  and the SC-FDE
scheme  can only deal with  the responses within the CP while the  responses beyond CP-length are regarded as
interference. Notice that the power gains for above-mentioned schemes are all optimized by their own algorithms.
 %so that the   results reflect their  best performance over different power gain factors.
A  half-duplex (HD)  relay scheme with frequency division duplex (FDD) is also compared. Considering the fairness of spectral efficiency, the  modulation mode of the  FDD scheme  is 16QAM.

\begin{figure}
\hspace{3cm}
%\vspace{-0.3cm}
\centering
\includegraphics[scale=0.32]{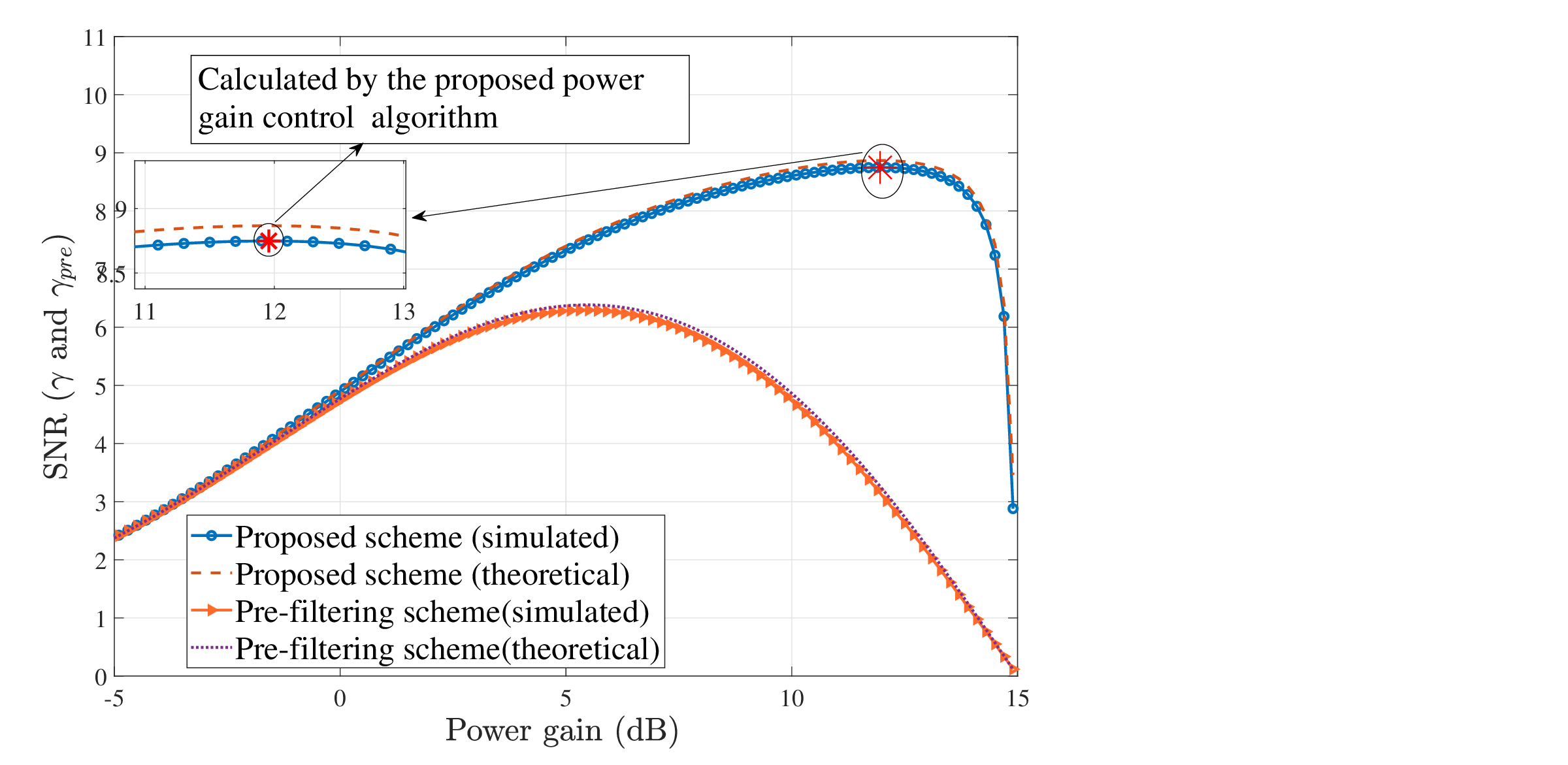}
\caption{SNR ($\gamma $ and $\gamma_{pre} $) performances with different power gains.} \label{fig:SNRinD}
\end{figure}

First we present a  simulation with  fixed channel coefficients to show the  SNR ($\gamma $ and $\gamma_{pre} $) performance with various  power gains when  $SNR_c\!\!=\!\!10$dB, which is shown in Fig.~\ref{fig:SNRinD}. The theoretical curves of the proposed scheme and the straightforward  pre-filtering scheme are plotted based on (\ref{eq1823}) and (\ref{eq1824}), respectively. After simulating an  OFDM transmission process, the  simulation results are  obtained by  separately calculating the noise power and the power of  useful signals. The theoretical results   are  concordant with the  simulated results. We can see the  proposed scheme  outperforms the
pre-filtering scheme for any power gain factors. The red star at the top  of  solid blue line is  the power gain factor calculated by the proposed power gain control algorithm.
%The simulation results  in Fig.~\ref{fig:snrbera} show that the  proposed scheme  outperforms the traditional  OFDM scheme  with standard CP structure  for any power gain factors.
%One can see that the rising of power gain at the relay, the  transmission performance first gets improved. However, too much power gain at the relay   leads to the deterioration in  BER  performance. The red star at the bottom of  solid line is  the power gain factor calculated by the proposed power gain control algorithm.
This simulation result verifies  the  necessity and effectiveness of the  power gain control algorithm.

\begin{figure}
%\hspace{-0.3cm}
\hspace{3cm}
%\vspace{-0.3cm}
%\setlength{\abovecaptionskip}{0.1cm}   % 调整图片标题与图距离
\centering
\includegraphics[scale=0.32]{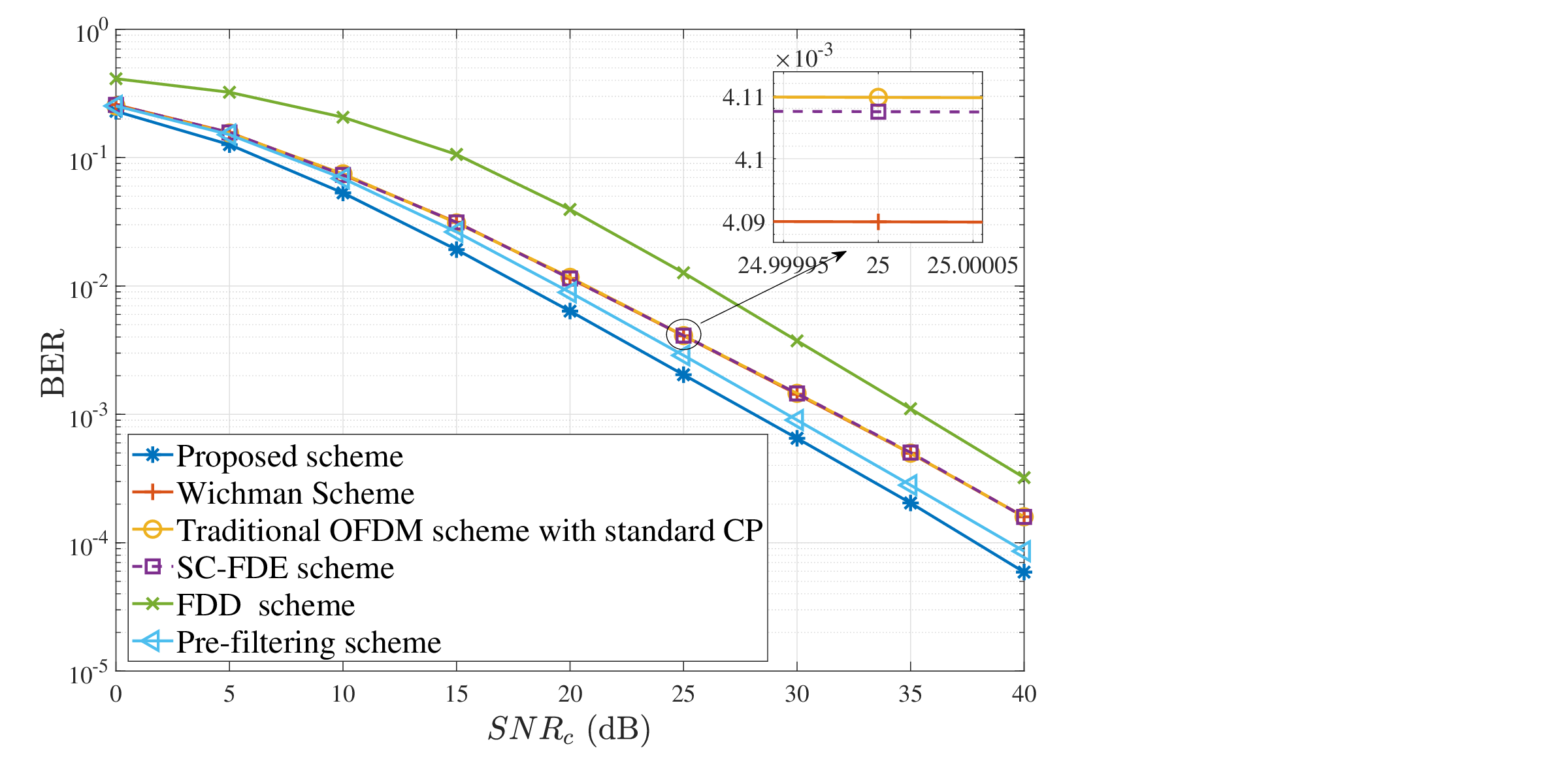}
\caption{BER performances without direct link for different schemes.} \label{fig:snrberb}
%\vspace{-0.5cm}
%\setlength{\belowcaptionskip}{0cm}
\end{figure}

\begin{figure}
%\hspace{-0.3cm}
\hspace{3cm}
%\vspace{-0.5cm}
%\setlength{\abovecaptionskip}{0.3cm}
\centering
\includegraphics[scale=0.32]{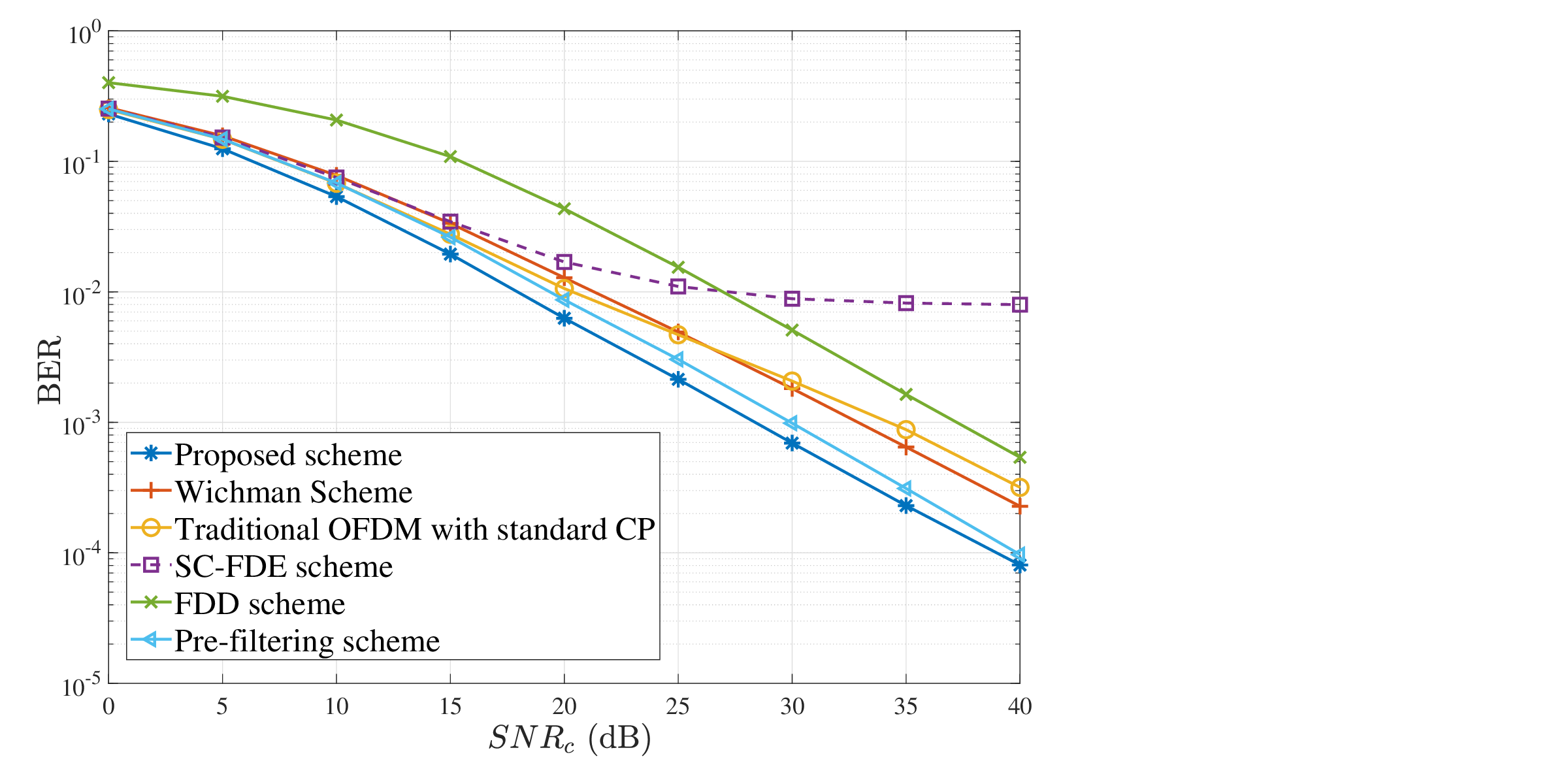}
\caption{BER performances with direct link for different schemes.} \label{fig:snrberc}
\end{figure}

From the BER  results without the direct link shown in Fig.~\ref{fig:snrberb}, consistent with the SNR analysis in Section~\ref{sec:Antenna}, the proposed scheme has better BER performance compared with the  straightforward  pre-filtering scheme and the other schemes.
%
%
% the proposed scheme outperforms the  other  schemes.
%Its performance gain compared with the traditional OFDM scheme is about 5dB.
%The other schemes all suffer  serious ISI because of the short CP length.
In Fig.~\ref{fig:snrberc}, considering  the direct link, the path loss of  $h_{sd}$ is set as 10dB. The results  in Fig.~\ref{fig:snrberc} indicate that the  power gain control algorithm  in Section~\ref{sec:Antenna}   still works and the proposed scheme outperforms the  other schemes,  when the direct link is not too strong.
Because the direct link  is not taken into consideration in~\cite{Yi2017SC13},
%regarded as interference  in  SC-FDE scheme,
% not taken into consideration
%
%Because the direct link
there exists an  error floor  for the SC-FDE scheme when $SNR_c>20$dB.
%its BER performance shows no improvement with the increase of $SNR_c$ when the $SNR_c$ is high.
 As for the proposed scheme, the  increased  improvement of BER performance  with $SNR_c$ increase  indicates that the direct link is treated as a  cooperative  signal  rather than an interference.

\begin{figure}
%\vspace{-0.4cm}
%\setlength{\abovecaptionskip}{0.03cm}
\hspace{3cm}
\centering
\includegraphics[scale=0.32]{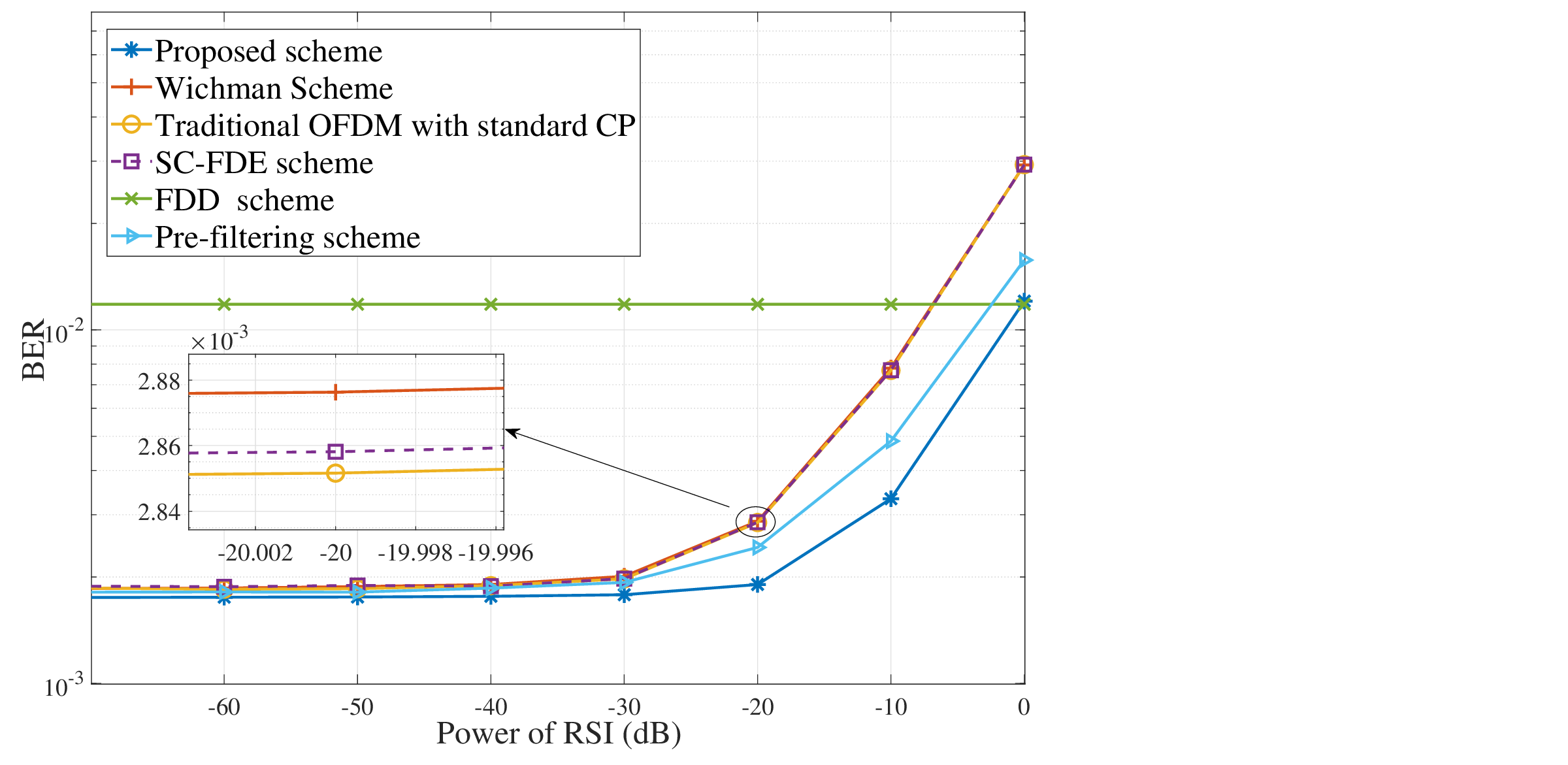}
\caption{BER performances with different RSI level.} \label{fig:snrberd}
%\vspace{-0.4cm}
\end{figure}

Lastly, without the direct link, we investigate the BER performance with the  RSI power when $SNR_c\!\!=\!\!25$dB in Fig.~\ref{fig:snrberd}.
 %When the RSI signals are weak, all the FD schemes can  outperform the HD scheme. But only
 Compared with the other FD schemes,  the proposed system  shows better  performance   under severe RSI  and thus  the strict performance requirement  of the SIC at relay may  be  relaxed.
%has the best performance
%  But with  the RSI level getting higher, only the proposed scheme can show a better performance than the HD scheme.
%The reason  for these results is that the proposed system  can collect  almost all the energy of the IIR channel responses. As for the other FD schemes, the energy of IIR channel responses within the CP-length are regarded as useful signals while the tail responses beyond CP-length are regarded as interferences.
 %The performance gap is more significant especially when the  SIC ability is insufficient.

%\vspace{-0.1cm}
\section{Conclusion}\label{sec:con6}
In this paper, based on  an equivalent IIR model for the FDR, following the newly proposed OFDM systems for IIR channels in~\cite{ANewxia},  we have presented  a  joint system design including  precoding, relay power gain control and equalization for  OFDM systems.
%The  GI at the source is  specially designed so that  the received OFDM block at the destination  has a CP structure and the same frequency domain equalization as the conventional OFDM systems applies.
%A power gain control algorithm    based on maximum SNR is introduced.
The simulation results show that the proposed scheme can achieve better  BER performance compared with the existing schemes.
 %The proposed system can  still  work well under severe RSI  and thus  the strict performance requirement  of the SIC at relay may  be  relaxed.

As a remark, in this letter we have only considered the case when all the point-to-point channels are flat fading, i.e., single path, for convenience. The idea proposed in this letter may be applied to broadband multipath channels, which is under our current investigation with more detailed analysis.

\bibliographystyle{IEEEtran}
\bibliography{IEEEabrv,References}

\end{document}